\documentclass[10pt,a4paper,twoside]{article}
\usepackage{epsfig}
\usepackage{baltlat6}
\usepackage{array}
\usepackage{here}
\begin{document}
\ \ \vspace{0.5mm} \setcounter{page}{300}

\titlehead{Baltic Astronomy, vol.\,99, 999--999, 2014}

 \titleb
 {DETERMINATION OF THE SOLAR GALACTOCENTRIC \\ DISTANCE FROM MASERS KINEMATICS}

\begin{authorl}
\authorb{A.T. Bajkova}{1} and
\authorb{V.V. Bobylev}{1,2}
\end{authorl}

\begin{addressl}
 \addressb{1}{Central (Pulkovo) Astronomical Observatory of RAS, 65/1 Pulkovskoye Ch.,
 St. Petersburg, Russia; anisabajkova@rambler.ru}
 \addressb{2}{Sobolev Astronomical Institute,
 St. Petersburg State University, Bibliotechnaya pl.2,
 St. Petersburg, Russia; vbobylev@gao.spb.ru}
\end{addressl}

\submitb{Received: 2014 December 99; accepted: 2014 December 99}

\begin{summary}
We have determined the Galactic rotation parameters and the solar
Galactocentric distance $R_0$ by simultaneously solving
Bottlinger's kinematic equations using data on masers with known
line-of-sight velocities and highly accurate trigonometric
parallaxes and proper motions measured by VLBI. Our sample
includes 93 masers spanning the range of Galactocentric distances
$R$ from 3 to 15 kpc. The solutions found are
 $\Omega_0 = 29.7\pm0.5$~km s$^{-1}$ kpc$^{-1},$
 $\Omega'_0 = -4.20\pm0.11$~km s$^{-1}$ kpc$^{-2}$,
 $\Omega''_0 =0.730\pm0.029$~km s$^{-1}$ kpc$^{-3},$ and
 $R_0=8.03\pm0.12$ kpc.
In this case, the linear rotation velocity at the solar distance
$R_0$ is $V_0=238\pm6$~km s$^{-1}$.
 \end{summary}


 \resthead
 {Determination of the Solar Galactocentric Distance}
 {A.T. Bajkova, V.V. Bobylev}

\sectionb{1}{INTRODUCTION}

Both kinematic and geometric characteristics are important for
studying the Galaxy, with the solar Galactocentric distance $R_0$
being the most important among them. Various data are used to
determine the Galactic rotation parameters. These include the line
of sight velocities of neutral and ionized hydrogen clouds with
their distances estimated by the tangential point method (Clemens
1985; McClure-Griffiths \& Dickey 2007; Levine et al. 2008),
Cepheids with the distance scale based on the period--luminosity
relation, open star clusters and OB associations with photometric
distances (Mishurov \& Zenina 1999; Rastorguev et al. 1999;
Zabolotskikh et al. 2002; Bobylev et al. 2008; Mel'nik and Dambis
2009), and masers with their trigonometric parallaxes measured by
VLBI (Reid et al. 2009a; McMillan \& Binney 2010; Bobylev \&
Bajkova 2010; Bajkova \& Bobylev 2012).

The solar Galactocentric distance $R_0$ is often assumed to be
known in a kinematic analysis of data, because not all of the
kinematic data allow $R_0$ to be reliably estimated. In turn,
different (including direct) methods of analysis give different
values of $R_0$.

Reid (1993) published a review of the $R_0$ measurements made by
then by various methods. He divided all measurements into primary,
secondary, and indirect ones and obtained the ``best value'' as a
weighted mean of the published measurements over a period of 20
years: $R_0=8.0\pm0.5$~kpc. Nikiforov (2004) proposed a more
complete three-dimensional classification in which the type of
$R_0$ determination method, the method of finding the reference
distances, and the type of reference objects are taken into
account. Taking into account the main types of errors and
correlations associated with the classes of measurements, he
obtained the ``best value'' $R_0=7.9\pm0.2$~kpc by analyzing the
results of various authors published between 1974 and 2003.

Based on 52 results published between 1992 and 2010, Foster \&
Cooper (2010) obtained the mean $R_0=8.0\pm0.4$~kpc.  Francis \&
Anderson (2013) gave a summary of 135 publications devoted to the
$R_0$ determination between 1918 and 2013. They concluded that the
results obtained after 2000 give a mean value of $R_0$ close to
8.0~kpc.

We have some experience of determining $R_0$ by simultaneously
solving Bottlinger's kinematic equations with the Galactic
rotation parameters. To this end, we used data on open star
clusters (Bobylev et al. 2007) distributed within about 4 kpc of
the Sun. Clearly, using masers belonging to regions of active star
formation and distributed in a much wider region of the Galaxy for
this purpose is of great interest. However, the first such
analysis for a sample of 18 masers performed by McMillan \& Binney
(2010) showed the probable value of $R_0$ to be within a fairly
wide range, 6.7--8.9~kpc. At present, the number of masers with
measured trigonometric parallaxes has increased (Reid et al.
2014), which must lead to a significant narrowing of this range.

The goal of this paper is to determine the Galactic rotation
parameters and the distance $R_0$ using data on masers with
measured trigonometric parallaxes.

\sectionb{2}{METHOD}
Here, we use a rectangular Galactic coordinate system with the
axes directed away from the observer toward the Galactic center
$(l$=$0^\circ$, $b$=$0^\circ,$ the $X$ axis), in the direction of
Galactic rotation ($(l$=$90^\circ$, $b$=$0^\circ,$ the $Y$ axis),
and toward the north Galactic pole ($b=90^\circ,$ the $Z$ axis).

The method of determining the kinematic parameters consists in
minimizing a quadratic functional $F:$
  \begin{equation}
   \begin{array}{lll}
 \min~F= \\
 \sum_{j=1}^N [w_r^j (V_r^j-\hat{V}_{r}^j)]^2+
 \sum_{j=1}^N [w_l^j (V_l^j-\hat{V}_{l}^j)]^2+
 \sum_{j=1}^N [w_b^j (V_b^j-\hat{V}_{b}^j)]^2
  \label{Functional}
 \end{array}
\end{equation}
provided that the following constraints derived from Bottlinger's
formulas with an expansion of the angular velocity of Galactic
rotation $\Omega$ into a series to terms of the second order of
smallness with respect to $r/R_0$ are fulfilled:
 \begin{equation}
 \begin{array}{lll}
 V_r=-u_\odot\cos b\cos l-v_\odot\cos b\sin l-w_\odot\sin b\\
 +R_0(R-R_0)\sin l\cos b \Omega'_0+0.5R_0 (R-R_0)^2 \sin l\cos b \Omega''_0,
\label{EQ-1}
 \end{array}
 \end{equation}
 \begin{equation}
 \begin{array}{lll}
 V_l= u_\odot\sin l-v_\odot\cos l+(R-R_0)(R_0\cos l-r\cos b) \Omega'_0\\
  +(R-R_0)^2 (R_0\cos l - r\cos b)0.5\Omega''_0 - r \Omega_0 \cos b,
 \label{EQ-2}
 \end{array}
 \end{equation}
 \begin{equation}
 \begin{array}{lll}
 V_b=u_\odot\cos l \sin b+v_\odot\sin l \sin b-w_\odot\cos b\\
    -R_0(R-R_0)\sin l\sin b\Omega'_0-0.5R_0(R-R_0)^2\sin l\sin b\Omega''_0,
\label{EQ-3}
 \end{array}
 \end{equation}
where $N$ is the number of objects used; $j$ is the current object
number; $V_r$ and $V_l,$ $V_b$ are the model values of the
three-dimensional velocity field: the line-of-sight velocity and
the proper motion velocity components in the $l$ and $b$
directions, respectively; $V_l=4.74 r \mu_l\cos b$, $V_b=4.74 r
\mu_b$ are the measured components of the velocity field (data),
with $\hat{V}_{r}^j, \hat{V}_{l}^j$ and $\hat{V}_{b}^j$, where the
coefficient 4.74 is the quotient of the number of kilometers in an
astronomical unit and the number of seconds in a tropical year;
$w_r^j, w_l^j,w_b^j$ are the weight factors; $r$ is the
heliocentric distance of the star calculated via the measured
parallax $\pi,$ $r=1/\pi$; the star's proper motion components
$\mu_l\cos b$ and $\mu_b$ are in mas yr$^{-1}$ (milliarcseconds
per year), the line-of-sight velocity $V_r$ is in km s$^{-1}$;
$u_\odot,v_\odot,w_\odot$ are the stellar group velocity
components relative to the Sun taken with the opposite sign (the
velocity $u$ is directed toward the Galactic center, $v$ is in the
direction of Galactic rotation, $w$ is directed to the north
Galactic pole); $R_0$ is the Galactocentric distance of the Sun;
$R$ is the distance from the star to the Galactic rotation axis,
 \begin{equation}
 \begin{array}{lll}
 R^2=r^2\cos^2 b-2R_0 r\cos b\cos l+R^2_0.
 \label{RR}
 \end{array}
 \end{equation}
$\Omega_0$ is the angular velocity of rotation at the distance
$R_0;$ the parameters $\Omega'_0$ and $\Omega''_0$ are the first
and second derivatives of the angular velocity with respect to
$R,$ respectively.

\begin{figure}[!tH]
 \label{delta}
\vbox{
\centerline{\psfig{figure=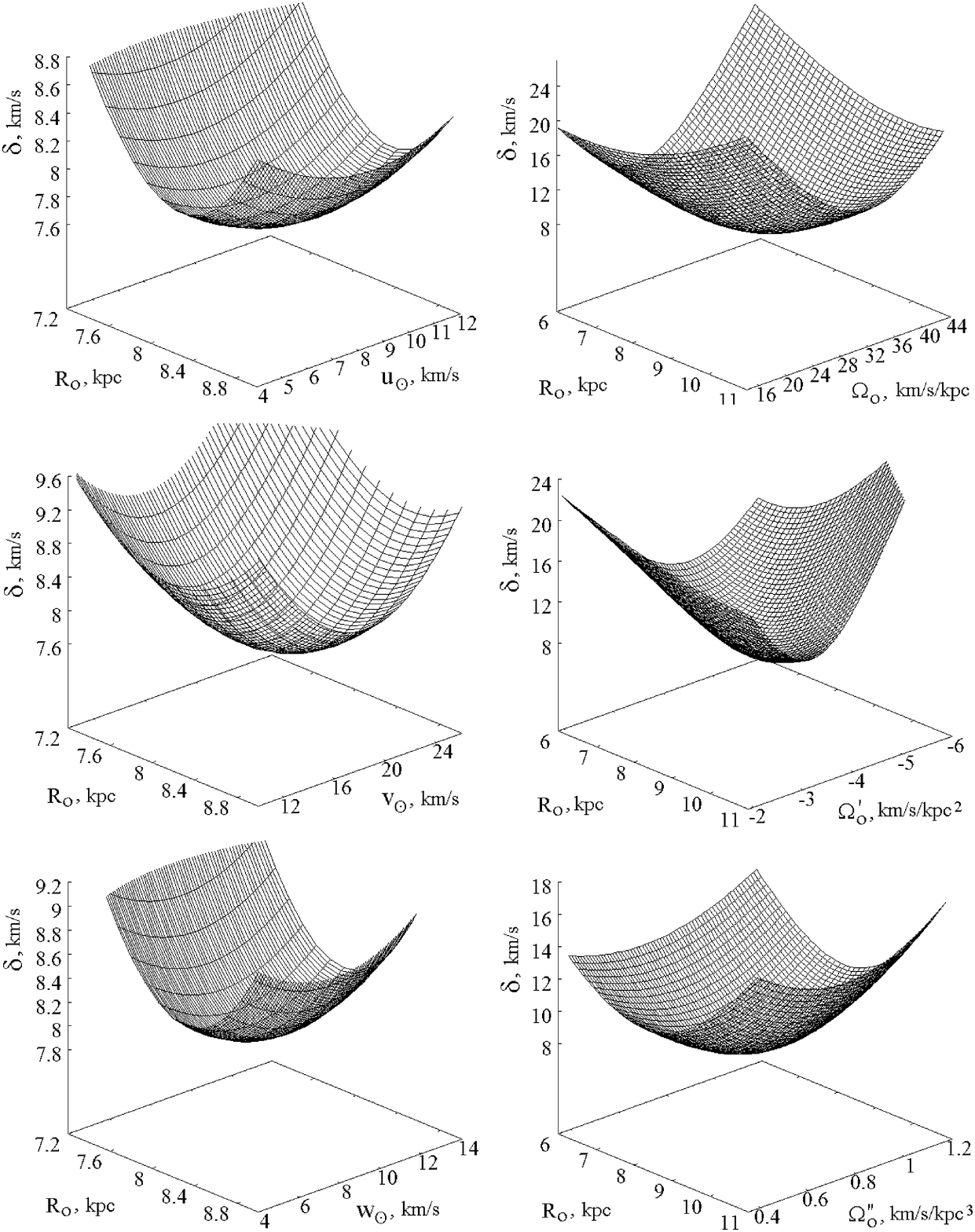,width=100mm,angle=0,clip=}}
\vspace{1mm} \captionb{1} {Graphical representation of the
two-dimensional residuals $\delta=\sqrt{F}$ corresponding to
solution (9); one of the measurements is specified by the
parameter $R_0$; one of the parameters $u_\odot, v_\odot, w_\odot,
\Omega_0, \Omega'_0,$ and $\Omega''_0$ acts as the second
measurement; the remaining parameters are fixed at the level of
the solutions obtained.} }
\end{figure}

The weight factors in functional (1) are assigned according to the
following expressions (for simplification, we omitted the index
$i$):
 \begin{equation}
 \begin{array}{rll}
 w_r=       S_0/\sqrt {S_0^2+\sigma^2_{V_r}},\quad
 w_l=\beta  S_0/\sqrt {S_0^2+\sigma^2_{V_l}},\quad
 w_b=\gamma S_0/\sqrt {S_0^2+\sigma^2_{V_b}},
 \label{WESA}
 \end{array}
 \end{equation}
where $S_0$ denotes the dispersion averaged over all observations,
which has the meaning of a ``cosmic'' dispersion taken to be
8~km~s$^{-1}$;  $\beta=\sigma_{V_r}/\sigma_{V_l}$ and
$\gamma=\sigma_{V_r}/\sigma_{V_b}$ are the scale factors, where
$\sigma_{V_r},\sigma_{V_l}$ and $\sigma_{V_b}$ denote the velocity
dispersions along the line of sight, the Galactic longitude, and
the Galactic latitude, respectively. The system of weights (6) is
close to that from Mishurov \& Zenina (1999). We take
$\beta=\gamma=1$ according by Bobylev \& Bajkova~(2014).

The errors of the velocities $V_l$ and $V_b$ are calculated from
the formula
 \begin{equation}
 \sigma_{(V_l,V_b)}=4.74r\sqrt{\mu^2_{l,b}\Biggl({\sigma_r\over r}\Biggr)^2+\sigma^2_{\mu_{l,b}}}.
 \label{Errors}
 \end{equation}
The problem of optimizing functional (1), given Eqs. (2)--(4), is
solved numerically for the seven unknown parameters $u_\odot$,
 $v_\odot$, $w_\odot$, $\Omega_0$, $\Omega'_0$, $\Omega''_0$, and $R_0$
from a necessary condition for the existence of an extremum. A
sufficient condition for the existence of an extremum in a
particular domain is the positive definiteness of the Hessian
matrix composed of the elements $\{a_{i,j}\}=d^2F/dx_i dx_j$,
where $x_i (i=1,...,7)$ denote the sought-for parameters,
everywhere in this domain. We calculated the Hessian matrix in a
wide domain of parameters or, more specifically, $\pm50\%$ of the
nominal values of the parameters.

Our analysis of the Hessian matrix for both cases of weighting
showed its positive definiteness, suggesting the existence of a
global minimum in this domain and, as a consequence, the
uniqueness of the solution. As an example, Fig.~1 shows the
two-dimensional residuals, or the square root of the functional
$F,$ with one of the measurements being specified by the parameter
$R_0$ and one of the parameters $u_\odot, v_\odot, w_\odot,
\Omega_0, \Omega'_0,$ and $\Omega''_0$, acting as the second
measurement, provided that the remaining parameters from the
series are fixed at the level of the solution obtained. The
presented pictures clearly demonstrate a global minimum in a wide
domain of parameters. In the case of unit weight factors, the
Hessian matrix is also positively defined far beyond this domain.
However, as will be shown below, the adopted weighting allowed the
accuracy of the solutions obtained to be increased.

We estimated the errors of the sought-for parameters through Monte
Carlo simulations. The errors were estimated by performing 100
cycles of computations. For this number of cycles, the mean values
of the solutions virtually coincide with the solutions obtained
purely from the initial data, i.e., without adding any measurement
errors.

\sectionb{3}{DATA}
Based on published data, we gathered information about the
coordinates, line-of-sight velocities, proper motions, and
trigonometric parallaxes of Galactic masers measured by VLBI with
an error, on average, less than 10\%. These masers are associated
with very young objects, protostars of mostly high masses located
in regions of active star formation.

One of the projects to measure the trigonometric parallaxes and
proper motions is the Japanese VERA (VLBI Exploration of Radio
Astrometry) project devoted to the observations of H$_2$O masers
at 22.2 GHz (Hirota et al. 2007) and a number of SiO masers (which
are very few among young objects) at 43 GHz (Kim et al. 2008).

Methanol (CH$_3$OH, 6.7 and 12.2 GHz) and H$_2$O masers are
observed in the USA on VLBA (Reid et al. 2009a). Similar
observations are also being carried out within the framework of
the European VLBI network (Rygl et al. 2010), in which three
Russian antennas are involved: Svetloe, Zelenchukskaya, and
Badary. These two programs enter into the BeSSeL project\footnote
{http://www3.mpifr-bonn.mpg.de/staff/abrunthaler/BeSSeL/index.shtml}
(Bar and Spiral Structure Legacy Survey, Brunthaler et al. 2011).

Initial data on 103 masers was taken from Reid et al.~(2014).

\sectionb{4}{RESULTS}
Using the three-dimensional maser velocity field for sample of 103
masers for equations with seven unknowns and weights ~(6) we
obtained  the following solution:
 \begin{equation}
 \begin{array}{lll}
 (u_\odot,v_\odot,w_\odot)=(5.20,17.47,7.73)\pm(0.74,0.72,0.32)~\hbox{km s$^{-1}$}, \\
  \Omega_0= 29.74\pm0.45~\hbox{km s$^{-1}$ kpc$^{-1}$}, \\
  \Omega'_0= -4.20\pm0.11~\hbox{km s$^{-1}$ kpc$^{-2}$}, \\
  \Omega''_0=  0.730\pm0.029~\hbox{km s$^{-1}$ kpc$^{-3}$},\\
        R_0= 8.03\pm0.12~\hbox{kpc},  \\
   \sigma_0= 10.59~\hbox{km s$^{-1}$},  \\
    N_\star=93.
 \label{final solution}
 \end{array}
 \end{equation}
Note, that in this case, ten sources (G000.67-00.03,
G010.47+00.02, G010.62-00.38, G023.70-00.19, G025.70+00.04,
G027.36-00.16, G009.62+00.19, G012.02-00.03, G078.12+03.63,
G168.06+00.82) were rejected according to the 3$\sigma$ criterion.
From this solution the linear rotation velocity at the solar
distance $R_0$ is
 $V_0=238\pm6$ km s$^{-1}$ and the Oort constants
 $A=0.5R_0\Omega_0'$ and
 $B=\Omega_0+0.5R_0\Omega_0'$ are
 $A=-16.86\pm0.45$~km s$^{-1}$ kpc$^{-1}$ and
 $B= 12.88\pm0.63$~km s$^{-1}$ kpc$^{-1}$.

As a clear illustration of the uniqueness of the solution obtained
(i.e., the existence of a global minimum of the functional F in a
wide range of sought for parameters), Fig.~1 presents the
two-dimensional dependence of the residuals $\delta=\sqrt{F}$
(see~(1)) on $R_0$ and one of the parameters $u_\odot, v_\odot,
w_\odot, \Omega_0, \Omega'_0,$ and $\Omega''_0$, provided that the
remaining parameters are fixed at the level of
solutions~(\ref{final solution}).

Figure~2 presents the Galactic rotation curve constructed with
parameters~(\ref{final solution}) using the value of
$R_0=8.03$~kpc found; when calculating the boundaries of the
confidence region, we took into account the uncertainty in
estimating $R_0$ of 0.12 kpc.

We also obtained a few solutions satisfying various limitations on
data (see Table 1). The results shown might be of some interest.
In the second column of the table we have the solution, obtained
with the use of nearly all masers ($N=101$), only two masers
(G000.67-00.03, G010.47+00.02) with the most unreliable velocities
were rejected. In the third and fourth columns there are the
solutions obtained with the limitation on the star galactocentric
distance $R$ and with a limited precision of parallaxes,
$e_\pi/\pi$. The first solution has a significant error unit of
weight $\sigma_0,$ in other cases, this value is significantly
less. Therefore we consider the first solution as the most
unreliable.
In the last column we give the results obtained when neglecting 23
masers that were flagged as outliers by Reid et al. (2014). As it
is seen other three solutions do not have principal differences.
In our opinion, the solution~(\ref{final solution}) obtained from
maximum amount of initial masers ($N=103$) is of most interest.

\begin{table}[!t]
\begin{center}
\vbox{\footnotesize\tabcolsep=3pt
\parbox[c]{124mm}{\baselineskip=10pt
{\smallbf\ \ Table 1.}{\small\ Kinematic parameters found by using
the three-dimensional velocity field.\lstrut}}
      \label{t1}
\begin{tabular}{|l|r|r|r|r|r|}\hline
   Parameters                           &                &   $4<R<12$~kpc & $e_\pi/\pi<12\%$ & \\\hline
\hstrut
   $u_\odot,$    km s$^{-1}$            & $ 6.85\pm0.75$ & $ 5.65\pm0.72$ & $ 6.04\pm0.77$ & $ 7.83\pm0.79$ \\
   $v_\odot,$    km s$^{-1}$            & $14.31\pm0.65$ & $15.48\pm0.73$ & $14.23\pm0.81$ & $13.25\pm0.75$ \\
   $w_\odot,$    km s$^{-1}$            & $ 7.74\pm0.35$ & $ 8.45\pm0.38$ & $ 8.24\pm0.42$ & $ 9.18\pm0.43$ \\
 $\Omega_0,$     km s$^{-1}$ kpc$^{-1}$ & $29.55\pm0.45$ & $29.49\pm0.43$ & $29.76\pm0.48$ & $29.39\pm0.46$ \\
 $\Omega^{'}_0,$ km s$^{-1}$ kpc$^{-2}$ & $-3.86\pm0.08$ & $-4.36\pm0.11$ & $-4.05\pm0.12$ & $-3.76\pm0.10$ \\
$\Omega^{''}_0,$ km s$^{-1}$ kpc$^{-3}$ & $ 0.59\pm0.02$ & $ 0.95\pm0.05$ & $ 0.68\pm0.03$ & $ 0.56\pm0.02$ \\
        $R_0,$   kpc                    & $ 8.25\pm0.41$ & $ 7.84\pm0.13$ & $ 8.10\pm0.13$ & $ 8.46\pm0.12$ \\
   $\sigma_0,$   km s$^{-1}$            &         12.49  &         9.97   &          9.60  &  8.86 \\
     $N_\star$                          &           101  &           88   &           78   &    80 \\\hline
\end{tabular}
 }
\end{center}
\end{table}

\sectionb{5}{DISCUSSION}
The parameters of the Galactic rotation curve we found~(\ref{final
solution}) are in good agreement with the results of analyzing
such young Galactic disk objects as OB associations, $\Omega_0
=31\pm1$ km s$^{-1}$ kpc$^{-1}$ (Mel'nik \& Dambis 2009), blue
supergiants,
 $\Omega_0=29.6\pm1.6$~km s$^{-1}$ kpc$^{-1}$ and
 $\Omega'_0=-4.76\pm0.32$~km s$^{-1}$ kpc$^{-2}$ (Zabolotskikh et
al. 2002), or distant OB3 stars ($R_0=8$~kpc),
 $\Omega_0 =31.9\pm1.1$~km s$^{-1}$ kpc$^{-1}$,
 $\Omega^{'}_0 = -4.30\pm0.16$~km s$^{-1}$ kpc$^{-2}$ and
 $\Omega^{''}_0 = 1.05\pm0.35$~km s$^{-1}$ kpc$^{-3}$ (Bobylev \& Bajkova 2013).
The solution~(\ref{final solution}) is in good agreement with
$V_0=254\pm16$ km s$^{-1}$ at $R_0=8.4$~kpc (Reid et al. 2009a)
and $V_0=244\pm13$ km s$^{-1}$ for $R_0=8.2$~kpc (Bovy et al.
2009) determined from a sample of 18 masers. Note also the paper
by Irrgang et al. (2013), who proposed three Galactic potential
models constructed using data on hydrogen clouds and masers, with
the velocity $V_0$ having been found to be close to 240~km
s$^{-1}$ and $R_0\approx8.3$~kpc.

Individual independent methods give an estimate of $R_0$ with an
error of 10--15\%. Note several important isolated measurements.
Based on POPULATION-II Cepheids and RR Lyr stars belonging to the
bulge and using improved calibrations derived from Hipparcos data
and 2MASS photometry, Feast et al. (2008) obtained an estimate of
$R_0=7.64\pm0.21$~kpc. Having analyzed the orbits of stars moving
around a supermassive black hole at the Galactic center (the
method of dynamical parallaxes), Gillessen et al. (2009) obtained
an estimate of $R_0=8.33\pm0.35$~kpc. According to VLBI
measurements, the radio source Sqr A* has a proper motion relative
to extragalactic sources of $6.379\pm0.026$ mas yr$^{-1}$ (Reid \&
Brunthaler 2004); using this value, Sch\"onrich (2012) found
$R_0=8.27\pm0.29$~kpc and $V_0=238\pm9$ km s$^{-1}$. Two H$_2$O
maser sources (Sgr~B2), are in the immediate vicinity of the
Galactic center, where the radio source Sqr A* is located. Based
on their direct trigonometric VLBI measurements, Reid et al.
(2009b) obtained an estimate of $R_0=7.9^{+0.8}_{-0.7}$~kpc.

Using 73 masers Bobylev \& Bajkova (2014) obtained an kinematic
estimate of $R_0=8.3\pm0.2$~kpc and $V_0=241\pm7$~km s$^{-1}$.
Based on 80 maser sources, Reid et al. (2014) obtained an
kinematic estimate of $R_0=8.34\pm0.16$~kpc and $V_0=240\pm8$~km
s$^{-1}$. Thus, our kinematic estimate of $R_0=8.03\pm0.12$~kpc is
in good agreement with the known estimates and surpasses some of
them in accuracy.

\begin{figure}[!tH]
 \label{rotcurve}
\vbox{
\centerline{\psfig{figure=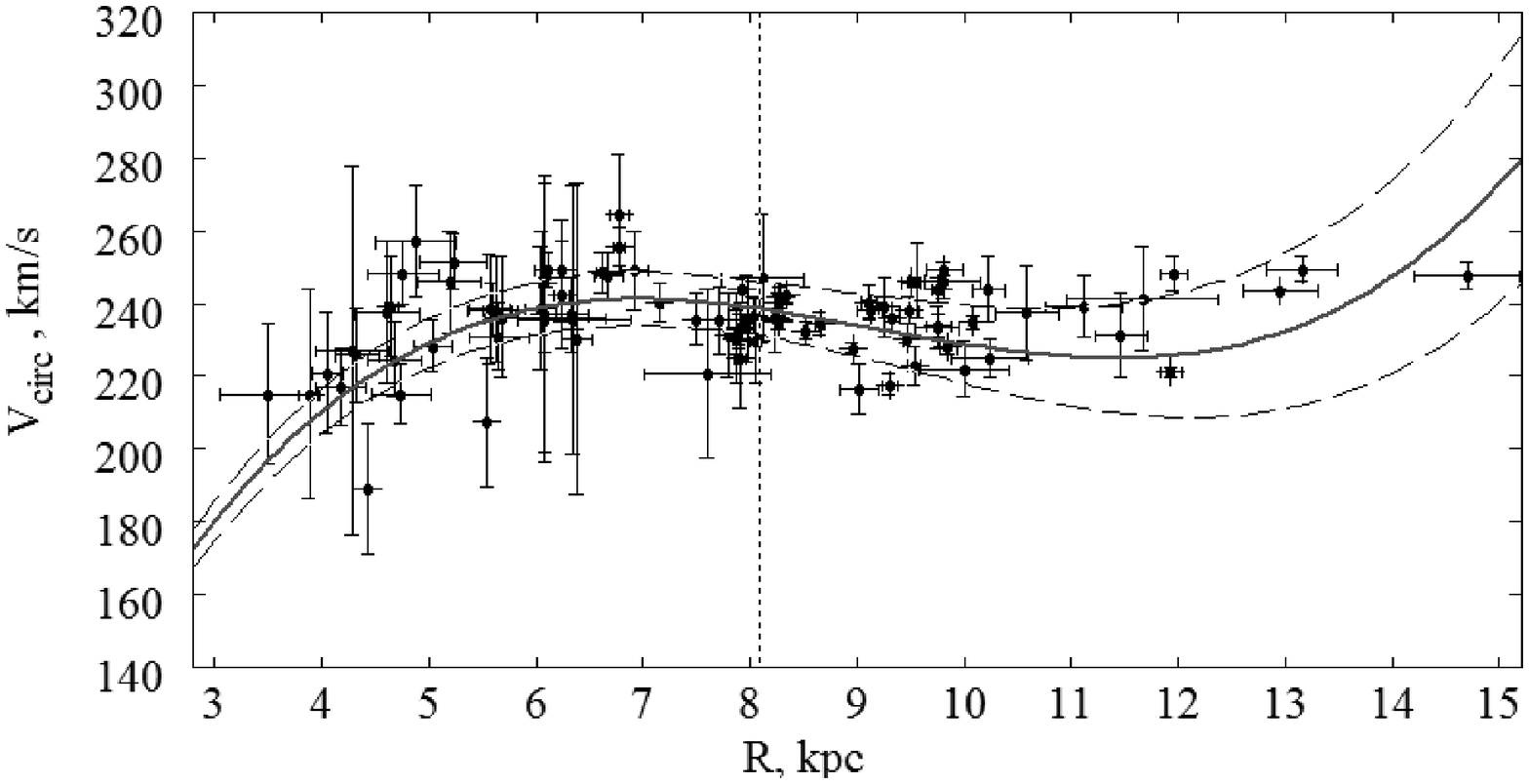,width=100mm,angle=0,clip=}}
\vspace{1mm} \captionb{2} {Galactic rotation curve constructed
with parameters~(\ref{final solution}) (thick line); the thin
lines mark the 1$\sigma$ confidence region; the vertical straight
line marks the Sun's position.} }
\end{figure}

\sectionb{6}{CONCLUSIONS}
Based on published data, we produced a sample of masers with known
line-of-sight velocities and highly accurate trigonometric
parallaxes and proper motions measured by VLBI. This allowed the
maser velocity field needed to solve Bottlinger's kinematic
equations to be formed. Bottlinger's kinematic equations we
considered relate the Galactic rotation parameters ($\Omega_0$ and
its derivatives), the solar Galactocentric distance $(R_0),$ the
object group velocity components relative to the Sun ($u_\odot,
v_\odot, w_\odot$). The method of minimizing the quadratic
functional that is the sum of the weighted squares of the
residuals of measurements and model velocities was used to find
the unknown parameters. Solutions were found in the cases of both
three-dimensional and two-dimensional velocity fields for various
numbers of sought-for parameters when various weighting methods
were applied. We established that the solution obtained from the
three-dimensional maser velocity field for seven sought-for
parameters ($u_\odot, v_\odot, w_\odot, \Omega_0, \Omega'_0,
\Omega''_0,$ and $R_0$) corresponding to the global minimum of the
functional in a wide range of their variations is most reliable.
This solution is~(\ref{final solution}). The linear Galactic
rotation velocity at the solar distance $R_0$ is $V_0=238\pm6$~km
s$^{-1}$. The solar Galactocentric distance $R_0$ is the most
important and debatable parameter. Our value $R_0=8.03\pm0.12$~kpc
is in good agreement with the most recent estimates and even
surpasses them in accuracy.

\thanks{
This work was
supported by the ``Nonstationary Phenomena in Objects of the
Universe'' Program P--21 of the Presidium of the Russian Academy
of Sciences.
 }

 \References
 \refb
 Bajkova A.T., Bobylev V.V. 2012, Astron. Lett. 38, 549

 \refb
 Bobylev V.V., Bajkova A.T., Lebedeva S.V. 2007, Astron. Lett. 33, 720

 \refb
 Bobylev V.V., Bajkova A.T., Stepanishchev A.S. 2008, Astron. Lett. 34, 515


 \refb
 Bobylev V.V., Bajkova A.T. 2010, MNRAS, 408,1788

 \refb
 Bobylev V.V., Bajkova A.T. 2013, Astron. Lett. 39, 532 

 \refb
 Bobylev V.V., Bajkova A.T. 2014, Astron. Lett. 40, 389 

 \refb
 Bovy J., Hogg D.W., Rix H.-W. 2009, ApJ, 704, 1704

 \refb
 Brunthaler A., Reid M.J., Menten K.M., et al. 2011, AN, 332, 461

 \refb
 Clemens D.P. 1985, ApJ, 295, 422




 \refb
 Feast M.W., Laney C.D., Kinman T.D., et al. 2008, MNRAS 386, 2115

 \refb
 Foster T., Cooper B. 2010, ASP Conf. Ser. 438, 16

 \refb
 Francis C., Anderson E. 2013, arXiv:1309.2629

 \refb
 Gillessen S., Eisenhauer F., Trippe S. 2009, et al., ApJ, 692, 1075

 \refb
 Hirota T., Bushimata T., Choi Y.K., et al. 2007, PASJ, 59, 897

 \refb
 Irrgang A., Wilcox B., Tucker E., et al. 2013, A\&A, 549, 137

 \refb
 Kim M.K., Hirota T., Honma M., et al. 2008, PASJ, 60, 991

 \refb
 Levine E.S., Heiles C., Blitz L. 2008, ApJ, 679, 1288

 \refb
 McClure-Griffiths N.M., Dickey J.M. 2007, ApJ, 671, 427

 \refb
 McMillan P.J., Binney J.J. 2010, MNRAS, 402, 934


 \refb
 Mel'nik A.M., Dambis A.K. 2009, MNRAS, 400, 518

 \refb
 Mishurov Yu.N., Zenina I.A. 1999, A\&A, 341, 81

 \refb
 Nikiforov I.I. 2004, ASP Conf. Ser. 316, 199

 \refb
 Rastorguev A.S., Glushkova E.V., Dambis A.K., et al. 1999, Astron. Lett. 25, 595

 \refb
 Reid M.J. 1993, Ann. Rev. Astron. Astrophys. 31, 345

 \refb
 Reid M.J., Brunthaler A. 2004, ApJ, 616, 872

 \refb
 Reid M.J., Menten K.M., Zheng X.W., et al. 2009a, ApJ, 700, 137

 \refb
 Reid M.J., Menten K.M., Zheng X.W., et al. 2009b, ApJ, 705, 1548

 \refb
 Reid M.J., Menten K.M., Brunthaler A., et al. 2014, ApJ, 783, 130 


 \refb
 Rygl K.L.J., Brunthaler A., Reid M.J., et al. 2010, A\&A, 511, A2


 \refb
 Sch\"onrich R. 2012, MNRAS, 427, 274



 \refb
 Zabolotskikh M.V., Rastorguev A.S., Dambis A.K. 2002, Astron. Lett. 28, 454


\end{document}